%% file: main.tex
\documentclass{article}

\usepackage{arxiv}

\usepackage[utf8]{inputenc} 
\usepackage[T1]{fontenc}    
\usepackage{hyperref}       
\usepackage{url}            
\usepackage{booktabs}       
\usepackage{amsfonts}       
\usepackage{nicefrac}       
\usepackage{microtype}      
\usepackage{lipsum}

\usepackage{amsmath}
\usepackage{caption}
\usepackage{subcaption}
\usepackage{xcolor}
\usepackage{graphicx}
\usepackage[superscript]{cite}
\usepackage{sidecap}
\usepackage{multirow}

\title{Collective Variables for Free Energy Surface Tailoring - Understanding and Modifying Functionality in Systems Dominated by Rare Events}

\author{
  Dan ~Mendels\thanks{danmendels@uchicago.edu}\\
  Pritzker School of Molecular Engineering\\
  University of Chicago\\
  Chicago, Illinois 60637, United States\\
  \And
  Juan J. de Pablo\thanks{depablo@uchicago.edu}\\
  Pritzker School of Molecular Engineering\\
  University of Chicago\\
  Chicago, Illinois 60637, United States\\
}

\begin{document}
\date{}
\maketitle

\begin{abstract}
\noindent 

We introduce a method for elucidating and modifying the functionality of systems dominated by rare events that relies on the automated tuning of their underlying free energy surface. The proposed approach seeks to construct collective variables (CVs) that encode the essential information regarding the rare events of the system of interest. The appropriate CVs are identified using Harmonic Linear Discriminant Analysis (HLDA), a machine-learning based method that is trained solely on data collected from short ordinary simulations in the relevant metastable states of the system. Utilizing the interpretable form of the resulting CVs, the critical interaction potentials that determine the system's rare transitions are identified and purposely modified to tailor the free energy surface in manner that alters functionality as desired. The applicability of the method is illustrated in the context of three different systems, thereby demonstrating that  thermodynamic and kinetic properties can be tractably modified with little to no prior knowledge or intuition.

\end{abstract}

\section*{Introduction}

There is considerable interest in using molecular models to design materials and molecules. That interest is partly a result of the greater availability of computational resources and important algorithmic advances. Furthermore, given the amounts of data molecular models and simulations generate, a new generation of artificial intelligence (AI) based algorithms could lead to significant advances in our ability to engineer new materials. 

A standard approach to the structure-function relationships that govern a system's behavior has been to collect data from large numbers of different realizations of the system, and to then apply machine learning and deep learning based algorithms to identify any discernible patterns. Such a strategy has largely been successful \cite{agrawal2016perspective,meredig2014combinatorial,yang2018microstructural,webb2020targeted,liu2018generative,mao2020designing,kim2020inverse,swanson2020deep,stokes2020deep,shen2021deep,gomez2018automatic}, but some of its weaknesses have also become apparent. These include (i) the volume of data that is required, (ii) the inability to interpret certain outcomes, given the black-box nature of many AI based algorithms, and (iii) the limited applicability beyond the underlying training domain. Overcoming these challenges is a matter of on-going research within the field of AI at large; here we propose an alternative approach that relies on computational molecular engineering to circumvent some of these limitations.  


Specifically, a method is proposed that relies on the notion that the functionality of a system can often be characterized using a dimensionally reduced representation of its free energy surface (FES), within a space spanned by a set of collective variables (CVs). Such a depiction serves to condense the most essential information about a system into a small number of parameters that can be tuned, through optimization algorithms such as that used in ref. \cite{long2018rational}, for design purposes. To the best our knowledge, the strategy adopted here represents a first step towards the use of CVs as a tool for rational materials design and engineering.

To explore the applicability of the proposed method, we focus on systems the functionality of which is associated with an underlying rough free energy landscape. Such systems reside primarily in metastable states, and undergo rare but important transitions between them. Examples of such transitions include nucleation, chemical reactions, and conformational changes in macromolecules. The behavior of the system depends on the relative stability of the metastable states, and the heights of the barriers that separate them. A systematic means of tailoring that relative free energy or the barriers' heights could therefore be used to target a specific, sought after functionality.

As indicated above, we achieve a reduced representation through a projection of the FES onto a set of CVs. Note that CVs are generally used in the context of enhanced sampling techniques, such as umbrella sampling \cite{Torrie1977}, metadynamics \cite{laio_parrinello_2002}, adaptive biasing force \cite{darve2001calculating} and, more recently, algorithms that rely on machine learning to construct a free energy surface \cite{guo2018adaptive,valsson_parrinello_2014,bonati2019neural}. In all these methods, more efficient sampling is attained by adding an external biasing potential in a form outlined by CVs constructed to correspond to the system's slowest modes. The addition of such an external bias thus effectively flattens the system's FES, thereby keeping it from lingering for too long in local minima. Generally speaking, CVs are functions of the system atomic coordinates $\mathbf{s}(\mathbf{R})$ and can be defined by the following relations:

\begin{equation}
\label{collective variables defined}
P(\mathbf{s})=\int{d\mathbf{R}\delta[\mathbf{s}-\mathbf{s(\mathbf{R})}}] P(\mathbf{R})
\end{equation}

\noindent where $P(\mathbf{s})$ is the system probability to possess a set  of given CV values $\mathbf{s}$, $P(\mathbf{R})$ is the Boltzmann probability, and $\delta$ is Kronecker's delta.
A system's FES with respect to the utilized set of CVs then follows:

\begin{equation}
\label{FES for CVs}
F(\mathbf{s})=-\frac{1}{\beta}\log P(\mathbf{s})
\end{equation}

\noindent where $\beta=1\slash k_BT$ and $k_B$ is Boltzmann's constant and $T$ is the temperature.

CVs that enable efficient sampling of phase space tend to encode the physical essence of the system's slow processes and functionality. This attribute makes them potentially useful tools for engineering. The construction of CVs can be challenging, however, and it has recently been shown that AI based methodologies can be of significant aid in this context \cite{ribeiro2018reweighted,mendels2018collective,bonati2020data,chen2018collective,wehmeyer2018time,noe2019boltzmann,sultan2018automated,mccarty2017variational}. Here, we employ the Harmonic Linear Discriminant Analysis (HLDA) \cite{mendels2018collective,piccini2018metadynamics,mendels2018folding,rizzi2019blind, zhang2019improving} method for its ease of use and straightforward interpretability. 

In HLDA, CVs consist of linear weighted sums of descriptors. The method requires as an input short trajectories obtained from simulations run in the system's various metastable states. Given their linear form, the interpretability of the HLDA CVs is straightforward; descriptors having larger weights (in absolute value) are considered to be associated with forces that hold higher physico-chemical importance with respect to the processes of interest.  The HLDA CV descriptors hierarchy can thus be used to identify the set of forces and interactions that should be tuned for the purpose of tailoring its FES and modifying a system's functionality in a desirable manner. 

We test the proposed approach, which we refer to as Collective Variables for Free Energy Surface Tailoring (CV-FEST), on three test cases of increasing complexity, namely: (a) a model molecule that undergoes conformational changes, b) a conformational change in a small peptide and c) the folding of a model protein. In all three cases, we find that tuning the interaction potentials of the system according to the proposed scheme enables tailoring of the FES, and the corresponding thermodynamic and/or kinetic properties in a simple, tractable manner.

\subsection*{Methods}

As noted above, we rely on HLDA to determine CVs \cite{mendels2018collective,piccini2018metadynamics}. HLDA requires as an input a list of user defined system descriptors $d_i$, e.g. distances between atoms, bond angles or more complex coordinates such as the enthalpy of a system or a surrogate to its entropy \cite{piaggi2017enhancing,Mendels2018,mendels2018collective}. With the current objective in mind, however, we limit here the types of descriptors that can be used to those that can be directly identified with tunable interaction potentials of the system, e.g. bonded and non-bonded interactions. In future work the utilization of more complex descriptors will also be explored. 

Once a descriptor set is assembled, HLDA requires calculation of the expectation value vectors $\mu_I$ and covariance matrices $\Sigma_I$ of the descriptor set for each of the metastable states $I\in{M}$ associated with the rare transitions under consideration. These can be estimated using data gathered in short unbiased simulation runs in each of the states. To construct the CVs, HLDA then estimates the directions $\mathbf{W}$ in the $N_d$ dimensional descriptor space on which the projections of the collected training distributions are best separated. This is done through the maximization of the ratio between the training data so-called between-class $\mathbf{S}_b$ and within-class $\mathbf{S}_w$ scatter matrices. The former is measured by the square of the distances between the projected means, and can be written as

\begin{equation}
\label{mean_transf_mat}
\mathbf{W}^T \mathbf{S}_b \mathbf{W}
\end{equation}

\noindent
with

\begin{equation}
\label{between_class}
\mathbf{S}_b = \left( \boldsymbol{\mu}_I - \boldsymbol{\bar{\mu}} \right)\left( \boldsymbol{\bar{\mu}} - \boldsymbol{\mu}_I \right)^T
\end{equation}

\noindent
where $\boldsymbol{\mu}_I$ represents the expectation value vector of the I-th metastable state and $\boldsymbol{\bar{\mu}}$ is the overall mean of the distributions, i.e. $\boldsymbol{\bar{\mu}}=1/M \sum_{I}^{M}\boldsymbol{\mu}_I$. The within-class matrix, in contrast, is estimated using the harmonic average of the metastable states' multivariate variances:  

\begin{equation}
\label{cov_transf_mat}
\mathbf{W}^T \mathbf{S}_w \mathbf{W}
\end{equation}

\noindent
where 

\begin{equation}
\label{harmonic_mean}
\mathbf{S}_w = \frac{1}{\frac{1}{\boldsymbol{\Sigma}_1} + \frac{1}{\boldsymbol{\Sigma}_2}+...+\frac{1}{\boldsymbol{\Sigma}_M}}.
\end{equation}

\noindent
Using the above terms, the HLDA objective function takes the form of a Rayleigh ratio
 
\begin{equation}
\label{fisher_ration}
\mathcal{J(\mathbf{W})} = \frac{\mathbf{W}^T \mathbf{S}_b \mathbf{W}}{\mathbf{W}^T \mathbf{S}_w \mathbf{W}}
\end{equation}

\noindent
which, given the normalization $\mathbf{W}^T \mathbf{S}_w \mathbf{W}=1$, can be shown to be equivalent to solving the eigenvalue equation \cite{piccini2018metadynamics}:

\begin{equation}
\label{maximizer}
\mathbf{S}_w^{-1}\mathbf{S}_b \mathbf{W}=\lambda\mathbf{W} 
\end{equation}

\noindent
Finally, the eigenvectors of Eq. \ref{maximizer} associated with the largest M-1 eigenvalues define the directions in descriptor space along which the distributions obtained from of the M metastable states least overlap, constituting the CVs that correspond to the rare transitions of interest. With these CVs at hand, we can now systematically tailor the system's FES and modify its functionality in a purposeful manner. To do this, the leading descriptors of each of the constructed CVs are identified and their corresponding interaction potentials are changed.  

\section*{Results}

\subsection*{Conformational change of a model molecule}

\begin{figure}
	\centering
    \includegraphics[trim=0 0 0 0,clip, width=0.96\columnwidth]{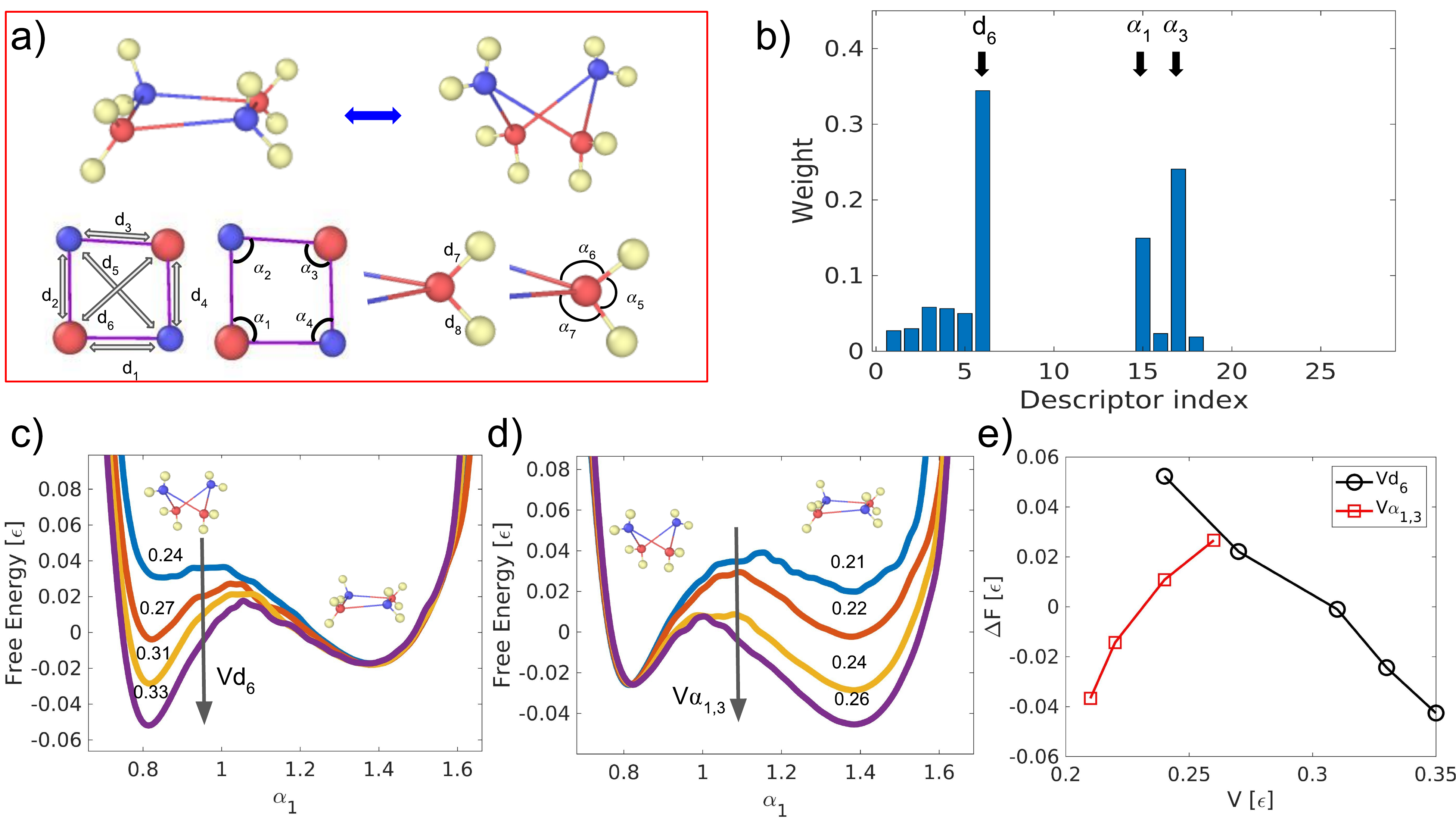}   
    \caption{(a) Illustration of the model molecule in its open and closed conformations and the descriptors used for the construction of the HLDA CV. (b) Absolute values of the HLDA CV weight distribution. (c) The molecule's FES as a function of the angle $\alpha_1$ for different values of the Lennard-Jones potential prefactor corresponding to the distance $d_6$. (d) The molecule's FES as function of the angle $\alpha_1$ for different values of the angular potential prefactor corresponding to angles $\alpha_1$ and $\alpha_3$. (d) Free energy difference between the open and closed states as a function of the modified interactions' strengths.
    }
    \label{fig:Cyclobutane}
\end{figure}

As a first test case we consider the conformational transition of a model molecule from an open conformation to a closed one (see Fig. \ref{fig:Cyclobutane}a). Our objective is to tune the relative stability of the conformations. To initiate CV-FEST, we start by assembling a list of 28 system descriptors, including bead to bead distances and bond angles. Next, we simulate two short trajectories in the two metastable conformations and calculate the collected distributions' expectation values and covariance matrices. Plugging these into Eq. \ref{maximizer} and solving, we obtain the HLDA CV corresponding to the conformational transition.

Fig. \ref{fig:Cyclobutane}b shows the absolute values of the  CV's weights. We denote by arrows the descriptors with the largest values, namely $d_6$, $\alpha_1$ and $\alpha_3$. Given the linear nature of the HLDA CV, we expect its weight distribution to only approximate the actual hierarchy of importance of the different interactions in the system. We thus focus our attention on the highest weighted descriptors. As will be shown in a subsequent example, improved accuracy in the ranking of the lower weighted descriptors can be attained by constructing new HLDA CVs over subsets of the original descriptor set, which exclude its highest ranking descriptors. 

Having identified the leading interactions of the conformational change, we next modify their strength. We focus on the thermodynamic attributes of the system reflected by the relative stability of the two metastable states, quantified by the free energy difference between them, $\Delta F$ (calculated using Eq. 9 in the SI). To determine for each interaction whether it should be strengthened or weakened for a given outcome, we examine the interaction potentials themselves (Fig. 4 in the SI). We find that while the minima of the force potentials $V\alpha_1$ and $V\alpha_3$ overlap with the open conformation state, the minimum of $Vd_6$ overlaps with the closed one. We thus deduce that for a given desired outcome, e.g. the stabilization of the open confirmation relative to the close one, $Vd_6$ needs to be modified in the opposite way to  $V\alpha_1$ and $V\alpha_3$.

To measure the FES of the molecule, we run long simulations in which numerous transitions between the two conformations occur, thereby arriving at a converged measure of the FES through the use of Eq. \ref{FES for CVs} (See Computational Details for more information). Figs. \ref{fig:Cyclobutane}c and \ref{fig:Cyclobutane}d show the molecule's FES obtained from simulations for different values of $Vd_6$, and $V\alpha_1$ and $V\alpha_3$, respectively. As can be seen, modifying the interaction strengths corresponding to the leading descriptors of the HLDA CV shifts the relative stability of the molecule's metastable states in a controlled way. Fig. \ref{fig:Cyclobutane}e summarizes these shifts by presenting the free energy differences between the states as a function of the modified interactions' strengths. While we have focused in this example on the molecule's thermodynamic profile, i.e. the relative stability of the states, we note that its kinetics, namely the transition rates between the states, can be tuned as well through the modification of the transition barrier. We have found that such a tuning can be achieved through combined modifications of $Vd_6$, and $V\alpha_1$ and $V\alpha_3$. For conciseness, however, we do not include the corresponding data here.

\subsection*{Alanine dipeptide}

\begin{figure}
	\centering
    \includegraphics[trim=0 0 200 0,clip, width=0.96\columnwidth]{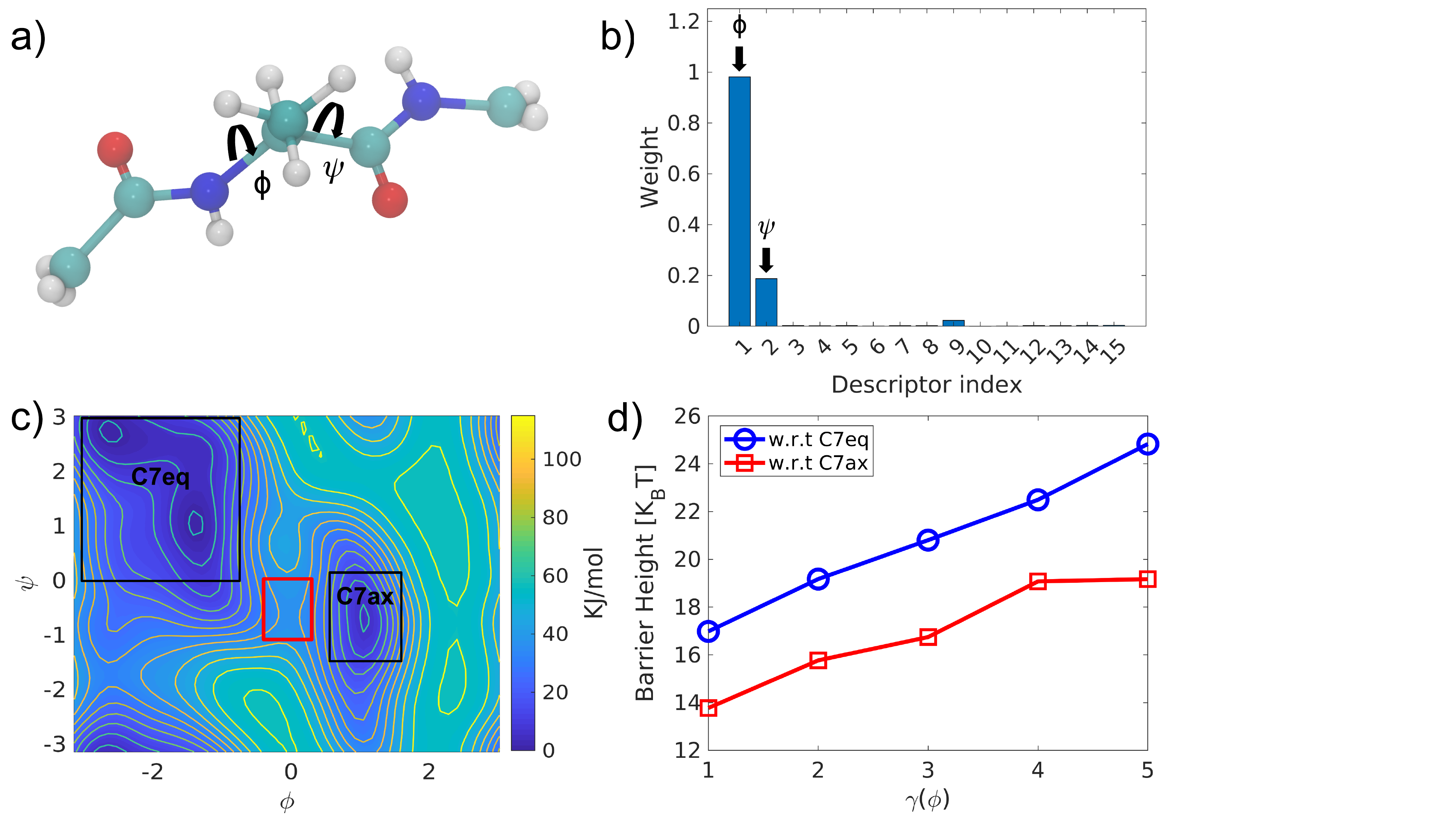}   
    \caption{(a) Illustration of alanine dipeptide \cite{HUMP96}. (b) Absolute values of the HLDA CV weights. For the descriptor index see the SI. (c) The FES of alanine dipeptide as a function of the dihedrals $\phi$ and $\psi$, calculated using WTMD. The transition region and metastable states are denoted by the red and black rectangles, respectively. (d) Free energy difference between the barrier region and the two metastable states, C7eq (Circles) and C7ax (squares), as a function of $\gamma (\phi)$, the factor by which the angle potential of $\phi$ was increased. 
    }
    \label{fig:Alanine}
\end{figure}

Our second test case involves the small peptide of alanine dipeptide in vacuum, frequently used as a test system for enhanced sampling targeted methods. The peptide can inhabit two conformations, the transition between which is a rare event characterized by a barrier height of $~17 k_BT$ at $T=300$K. We focus in this example on the tuning of the barrier height, given its importance in determining the system's kinetic behaviour, i.e. the rate at which the system undergoes transitions between states.  As before, we start by assembling a set of descriptors, potentially relevant to the conformational transition, opting for the set of 33 dihedral angles that have non-zero force potentials. We then run two unbiased simulations in the system's two metastable states to obtain their corresponding multivariate distributions in the space spanned by the descriptor set. To prevent the skewing of the HLDA output, prior to its use we calculate the correlation matrices of the distributions, and omit from the descriptor set descriptors that exhibit correlations greater than 0.9 to others in the set. Left with a set of 15 dihedral angles, we apply HLDA to obtain the CV corresponding to the conformational transition, as shown in Fig. \ref{fig:Alanine}b. As can be seen in the figure, we find that the CV is dominated by the $\phi$ and $\psi$ dihedrals (denoted in Fig. \ref{fig:Alanine}a) in accordance with the literature \cite{Barducci2008}. 

To alter the height of the transition barrier we modify the magnitude of the force potential corresponding to the $\phi$ dihedral angle. To measure the influence of the modifications on the barrier height, we run well tempered metadynamics simulations (WTMD) \cite{Barducci2008} using $\phi$ and $\psi$ as the biased CVs (See Computational Details for more information). Fig. \ref{fig:Alanine}c shows the FES obtained from such simulations. The transition region and metastable states are denoted by the red and black rectangles, respectively. Using Eq. 9 in the SI, we calculate the free energy differences between the barrier region and the metastable basins. Fig. \ref{fig:Alanine}d presents these differences as a function of $\gamma (\phi)$, the factor by which the angle potential corresponding to $\phi$ was strengthened. As can be seen, through these pinpointed modifications, the barrier height can be changed by up to $8 k_BT$, which corresponds to a change of more than three orders of magnitude in the rate of transitions between the metastable states.

\subsection*{Model protein}

\begin{figure}[!htbp]
	\centering
    \includegraphics[width=0.96\columnwidth]{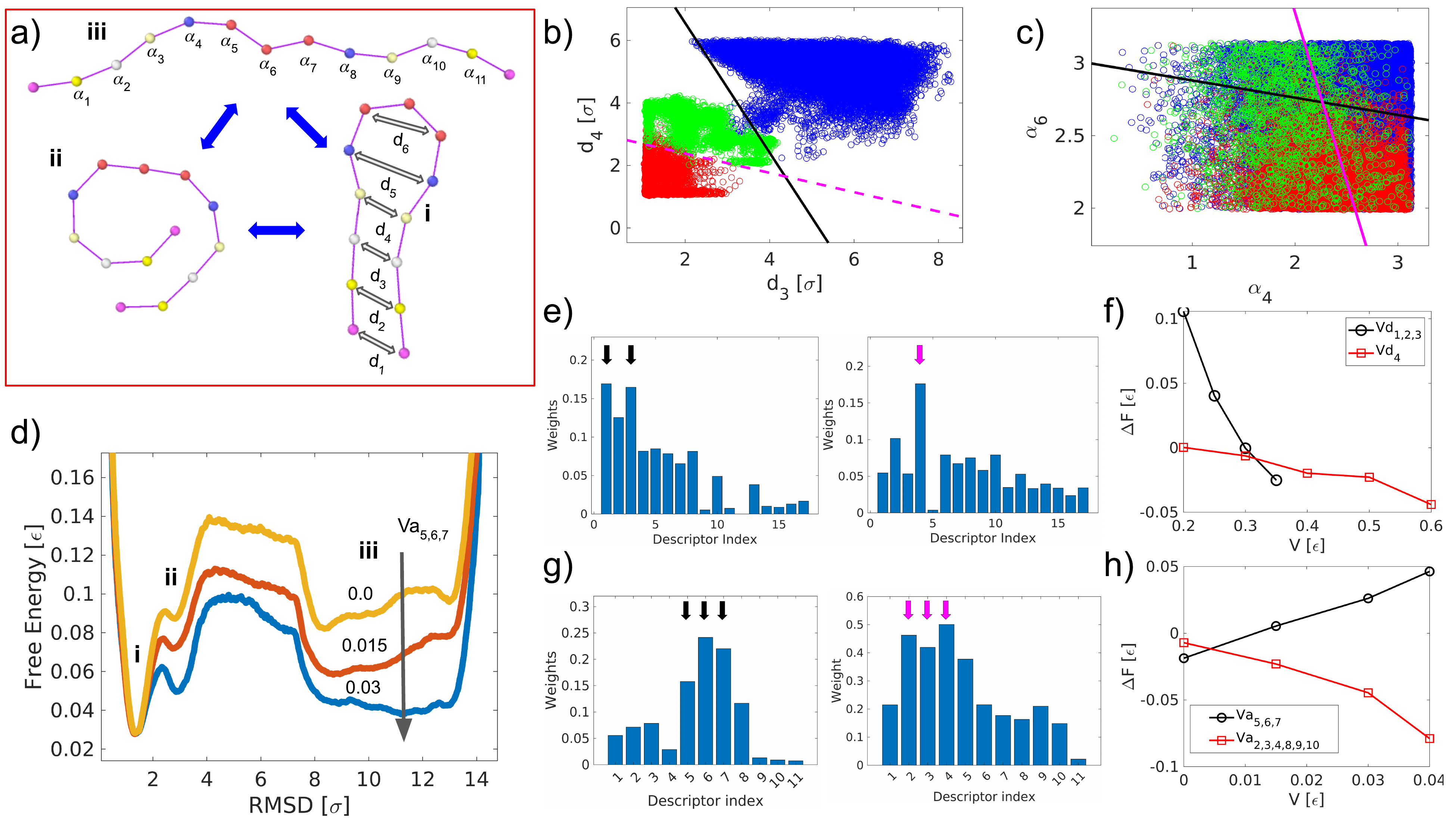}   
    \caption{(a) Model protein's unfolded, hairpin and toroid states. (b) Distributions collected from the protein hairpin (red), toroid (green) and unfolded (blue) states, and the \emph{full descriptor set} HLDA CV hyperplanes, projected on the plane spanned by the descriptors $d_3$ and $d_4$ (c) Distributions collected from the protein hairpin (red), toroid (green) and unfolded (blue) states, and the \emph{angle descriptor set} HLDA CV hyperplanes projected on the plane spanned by the descriptors $\alpha_4$ and $\alpha_6$ (d) FES of the protein as a function of the RMSD with respect to the hairpin state.  (e) Absolute values of the HLDA CV 1 (left) and CV 2 (right) weight distributions, both constructed from the full descriptor set. The arrows denote the selected highest weighted (in absolute value) descriptors of each of the CVs. Descriptors 1-6 correspond to the distances $d_1-d_6$ and descriptors 7-18 correspond to the angles $\alpha_1-\alpha_{11}$, respectively. (f) Free energy difference between the metastable states as a function of the magnitude of the interaction potentials corresponding to the highly weighted descriptors of the HLDA CVs constructed over the full set of descriptors. (g) Absolute values of the HLDA CV 1 (left) and CV 2 (right) weight distributions, both constructed over the angle descriptor set. The arrows indicate the selected highest weighted descriptors (in absolute value) of each of the CVs. Descriptors 1-11 correspond to the angles $\alpha_1-\alpha_{11}$, respectively. (h) Free energy difference between the metastable states as a function of the magnitude of the interaction potentials corresponding to the highly weighted descriptors of the HLDA CVs constructed over the angle set of descriptors. 
    }
    \label{fig:ToyProtein}
\end{figure}

As a final example we consider the folding of a model protein, utilizing the framework introduced in Ref. \cite{stillinger1993toy}. The protein consists of three metastable states (see Fig. \ref{fig:ToyProtein}a): an unfolded state, a hairpin and a toroid. Our objective in this case is to modulate the relative thermodynamic stability of the protein's three states. As in the previous examples, we start by assembling a set of descriptors that are potentially relevant to the transitions between the metastable states, including inter-bead distances and bond angles. Next, we run three short unbiased simulations in the protein's metastable states, and use the collected information to compute the expectation values and covariance matrices for each of the states. 

Given that the protein consists of three metastable states, two HLDA CVs are constructed, such that the hyperplanes to which they correspond separate the states from each other. Fig. \ref{fig:ToyProtein}e shows the absolute values of the weight distributions of the two resulting CVs, where \ref{fig:ToyProtein}b illustrates the CV hyperplanes projected along with the three metastable distributions obtained from the unbiased simulations on the plane spanned by the highly weighted descriptors $d_3$ and $d_4$. As can be seen in the figure, the hyperplane projections separate the three metastable states quite well within this particular plane.

Next, we set out to use the acquired information in order to vary the free energy differences between the system's different states.  Selecting the descriptors with highest weights in absolute value of the first and second HLDA CVs (denoted by arrows in Fig. \ref{fig:ToyProtein}e), we modify the magnitude of their corresponding interactions and calculate the influence on the system FES. To calculate the FES of the system we run long simulations in which multiple transitions between the three states can be observed. For convenience, we use the Root Mean Square Distance (RMSD) measure with respect to the system hairpin state as a CV upon which to project the system FES using Eq. \ref{FES for CVs} (See Computational Details for more information). The FESs measured in this way are shown in Fig. \ref{fig:ToyProtein}d. Three minima corresponding to the three metastable states of the system can be observed. Fig. \ref{fig:ToyProtein}d also illustrates how the FES is altered upon modifying the magnitude of the selected set of interaction potentials. To quantify the changes of the FES as function of the interaction potential modifications, we compute the free energy differences between the metastable states for each of the modified systems using Eq. 9 in the SI. Fig. \ref{fig:ToyProtein}f presents the outcomes of these calculations. As can be seen, modifying the magnitudes of the interactions corresponding to the highest weighted descriptors of the first HLDA CV shifts the free energy difference between the folded and unfolded states to the extent that the unfolded state becomes the system's global minimum. Similarly, modifying the magnitudes of the  interactions corresponding to the highest weighted descriptors of the second HLDA CV shifts the free energy difference between the two folded states.

So far, to tune the protein FES, we have relied only on modifying its non-bonded interactions.  To investigate if the system FES could also be tuned using CVs constructed only from the system's angles, we omitted the bead-to-bead distance descriptors from the descriptor set and reapplied HLDA on the remaining subset of descriptors. Fig. \ref{fig:ToyProtein}g exhibits the absolute values of the weight distributions of the two CVs, and Fig. \ref{fig:ToyProtein}c illustrates the projected CV hyperplanes and the three metastable distributions on the plane spanned by the highly weighted descriptors $\alpha_4$ and $\alpha_6$. While the separation between the distributions on the plane is not as definitive as that seen in Fig. \ref{fig:ToyProtein}b, we find that the CVs constructed in this case are also able to effectively tune the free energy differences between the metastable states.

As before, to tailor the system FES we selected the highest weighted descriptors in absolute value for each of the constructed CVs (denoted by arrows in \ref{fig:ToyProtein}g), and modified their corresponding interaction potential strengths. We note that given that in the construction of the HLDA CVs we relied on data collected from only one of the protein's toroid states (the second being a mirror reflection of the first), we appended to the highly weighted descriptors of the second CV ($\alpha_2$, $\alpha_3$ and $\alpha_4$) their symmetrical counterparts $\alpha_8$,$\alpha_9$ and $\alpha_{10}$, corresponding to the second toroid state. Fig. \ref{fig:ToyProtein}h displays the free energy difference between the metastable states as a function of the modified interaction parameters. One can see that the CVs constructed from the angle descriptor set are as effective as those constructed from the full set in terms of altering the free energy differences between metastable states. The possibility of constructing more than one set of CVs for a given system can provide further flexibility for tailoring its FES, and can potentially lead to more structural and chemical ways to changing it.

\section*{Computational Details}

\subsection*{Simulations of model molecule}

Simulations of the model molecule were run using LAMMPS \cite{PLIMPTON1995} and the PLUMED 2.4 plugin \cite{tribello_2014} utilizing reduced units, with $\epsilon=1$, $\sigma =1$, $m=1$ and $k_B=1$. Simulations were run in a three dimensional periodic box, where a constant temperature of $T=0.01$ was maintained using a Langevin thermostat \cite{schneider1978molecular} with a damping coefficient of 1 and a time step of 0.001. The interaction potentials of the pre-modified system were as follows: the angular potentials were harmonic, with angle stiffness and rest angle of $0.24$ and $\pi /2$, respectively for angles $\alpha_1-\alpha_4$, angle stiffness and rest angle of $0.12$ and $\pi /2$, respectively for angles $\alpha_5$, $\alpha_8$, $\alpha_{11}$ and $\alpha_{14}$, and angle stiffness and rest angle of $0.12$ and $\pi /2$, respectively for angles $\alpha_{6}$, $\alpha_{7}$, $\alpha_{9}$,$\alpha_{10}$, $\alpha_{12}$, $\alpha_{13}$ and $\alpha_{15}$, $\alpha_{16}$. Bond interactions were harmonic with a bond stiffness of $0.5$, rest length of $1$ and cutoff of $1.1$ for bonds $d_1-d_4$, a bond stiffness of 3, rest length of 0.2 and cutoff of 1.1 for bonds $d_5-d_{12}$. Non-bonded interactions were of the Lennard-Jones type, with $Vd= \epsilon^*=0.28$, $\sigma^*=0.7$ and $r_{cutoff}=2.5$ for $d_5$ and $d_6$.  In addition, harmonic restraints were placed on the angles $\alpha_2$ and $\alpha_4$ with a rigidity pre-factor of 500 to impede the formation of angles smaller than $1.2$, and on the angles $\alpha_5-\alpha_{20}$  with a rigidity pre-factor of 200 to restrict the formation of angles smaller than $0.72$.

\subsection*{Simulations of alanine dipeptide}

Simulations of alanine dipeptide (Ace-Ala-Nme) in vacuum were conducted using GROMACS 2018.4 \cite{berendsen1995gromacs,abraham2015gromacs} and the PLUMED 2.6 plugin \cite{tribello_2014}. The Amber99 \cite{hornak2006comparison} force field was employed.  A time step of 2 fs was used and a constant temperature of $300$K was maintained by the velocity rescaling thermostat of Bussi et al. \cite{bussi2007canonical}. All bonds involving hydrogen atoms were constrained with the linear constraint solver (LINCS) algorithm \cite{hess1997lincs}.  Electrostatic interactions were calculated with the particle mesh Ewald scheme \cite{essmann1995smooth} and a 1 nm cutoff was applied to all non-bonded interactions. 

WTMD simulations were run using the dihedral angles $\phi$ and $\psi$ as the CVs. The initial hill heights were taken to be 1 kJ/mol and hill widths were $\sigma_{phi}=0.35nm$ and $\sigma_{psi}=0.35nm$. The bias factor was 6 and the hill deposition rate was 500. 

\subsection*{Simulations of model protein}

Simulations of the model protein were run using LAMMPS \cite{PLIMPTON1995} and the PLUMED 2.4 plugin \cite{tribello_2014} utilizing reduced units, with $\epsilon=1$, $\sigma =1$, $m=1$ and $k_B=1$. Simulations were run in a two-dimensional periodic box, where a constant temperature of $T=0.02$ was maintained using a Langevin thermostat \cite{schneider1978molecular} with a damping coefficient of 1 and a time step of 0.001. All bond potentials were harmonic with a rest length of 1 and bond stiffness of 50. All angle potentials were harmonic as well with a rest angle of $\pi$ and an angle stiffness of 0.015 in the premodified protein. In addition, harmonic restraints with a stiffness pre-factor of 500 were placed on all angles to avoid the formation of angles smaller than $2$ radians. The interaction potentials between non-bonded beads followed a Lennard-Jones form, with  $Vd= \epsilon^* = 0.3$, $\sigma^* = 1$ and a cutoff of $r_{cutoff}=1.25$ for distances $d_1$, $d_2$, $d_3$ and $d_4$, with $Vd= \epsilon^* = 0.06$, $\sigma^* = 1$ and a cutoff of $r_{cutoff}=1.25$ for $d_5$, and with $Vd= \epsilon^* = 0.0$ for $d_6$. Lorentz-Berthelot mixing rules were applied for all non-bonded interactions.

\section*{Discussion and Conclusions}

A machine-learning based method has been presented to modify the functionality of systems dominated by rare events through the tailoring of their underlying FES. The method relies on the notion that the functionality of a system can be captured by a dimensionallity reduced representation of its FES within a space spanned by a set of suitably constructed collective variables (CVs). The applicability of the method was illustrated in the context of three systems whose behaviour is influenced by rare transitions. 

The proposed approach focuses on tuning the interaction potentials of considered systems in a selective manner, a procedure that can be translated to concrete physico-chemical modifications when real systems are considered. In biological systems for example, this may entail the implementation of point mutations, as done in ref. \cite{gao1989hidden}, whereas in non-biological supra-molecular systems, one might recourse to modifying or exchanging functional groups. We anticipate the method to be especially useful for large, complex systems that are commonly simulated using atomistic or coarse-grained force fields \cite{perego2020multiscale,bochicchio2019defects,cohen2021anisotropic,alessandri2017bulk}, where there is generally greater flexibility in translating the tuning of interaction potentials to real physico-chemical modifications. In this regard, we also expect CV-FEST to be a useful tool for force field tuning or optimization. 

While we have focused in this letter on systems dominated by rare events for which training sets are composed of trajectories run in the metastable states of the system, we anticipate the method's utility to extend to cases in which training sets consist of trajectories run in unstable states as well, e.g. transition states \cite{brotzakis2019augmented}, or to systems, the functionality of which is not dominated by rare events per se. Finally, we reiterate two of the prominent advantages of the proposed method. First, it requires orders of magnitude less training data than commonly used AI-based methods for materials design. Second, the method is interpretable, and can therefore provide valuable insights regarding the structure-function relationships in the system of interest. Such insights, we believe, might allow for development of new design rules, applicable beyond the systems on which the method was originally applied.

\section*{Acknowledgements}
D.M. would like to thank Zoran Bjelobrk for helpful discussions and Dr. Phwey Gil and Dr. Riccardo Alessandri for carefully reading the manuscript.

\section*{Appendix: software}
CV-FEST and HLDA will be publicly available for reference at \\ https://github.com/SSAGESLabs/PySAGES.

\newpage
\bibliographystyle{unsrt} 
\bibliography{library}
\newpage
\input{SI.tex}

\end{document}

%% file: SI.tex
\begin{center}
\vspace{5mm}
\LARGE{\textsc{Supporting information}}
\vspace{2mm}
\end{center}

\subsection*{Equation for the calculation of the free energy difference $\Delta F$ between two states A and B:}

\begin{equation}
\label{DeltaF}
\Delta F=-\frac{1}{\beta} \log{\frac{\int_A d\mathbf{s} e^{-\beta F(\mathbf{s})}}{\int_B d\mathbf{s} e^{-\beta F(\mathbf{s})}}}
\end{equation}

\subsection*{Plots of interaction potentials associated with leading descriptors of model molecule:}
\begin{figure}[!htbp]
	\centering
    \includegraphics[trim=0 0 0 0,clip, width=0.96\columnwidth]{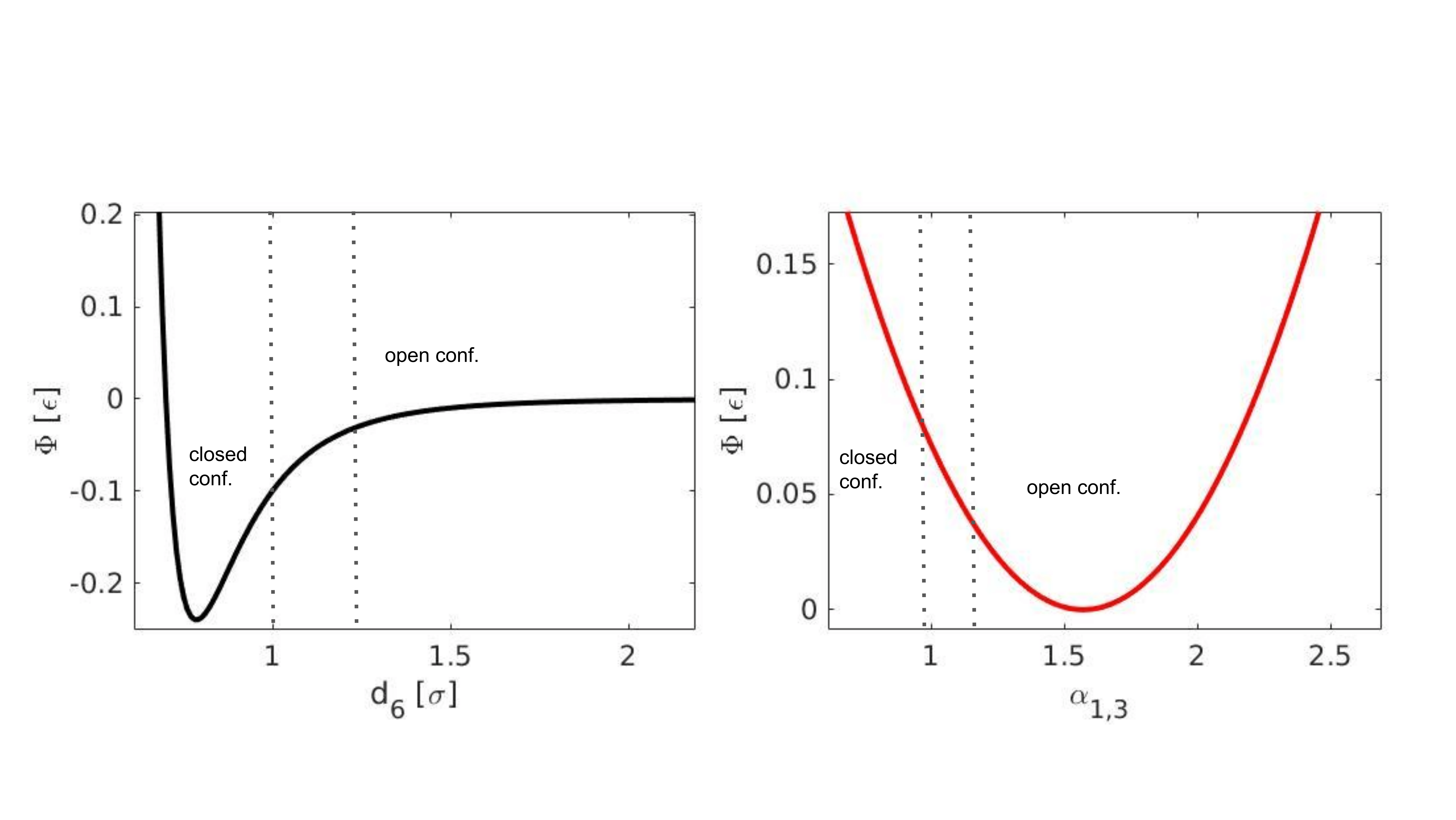}   
    \caption{Left: Lennard Jones interaction corresponding to the distance $d_6$. Right: angle potentials associated with the angles $\alpha_1$ and $\alpha_3$.  
    }
    \label{fig:Cyclobutane_interactions}
\end{figure}

\begin{table}[!htbp]
\caption{Index list (corresponding to Figure 2 in the main text) of alanine dipeptide descriptors (atom numbers in each dihedral angle specified.)}
\centering
\begin{tabular}{lcccccccc}
\toprule

Index                 & & atom 1     & atom 2  & atom 3    & atom 4       \\
\midrule
1         & &      5 &      7 &      9  &       15 &        \\
2         & &      7 &      9 &      15 &       17 &        \\
3         & &      1 &      2 &      5  &        6 &        \\
4         & &      1 &      2 &      5  &        7 &        \\
5         & &      3 &      2 &      5  &        6 &        \\
5         & &      3 &      2 &      5  &        7 &        \\  
6         & &      4 &      2 &      5  &        6 &        \\
8         & &      4 &      2 &      5  &        7 &        \\
9         & &      2 &      5 &      7  &        8 &        \\
10        & &      7 &      9 &      11 &       13 &        \\
11        & &      7 &      9 &      11 &       14 &        \\
12        & &      10 &     9 &      11 &       12 &        \\
13        & &      10 &     9 &      11 &       13 &        \\
14        & &      10 &     9 &      11 &       14 &        \\
15        & &      15 &     9 &      11 &       12 &        \\

\bottomrule
\end{tabular}
\label{tab:boxspecs}
\end{table}